# Photoelectronic mapping of spin-orbit interaction of intense light fields

Yiqi Fang,[1] Meng Han,[1] Peipei Ge,[1] Zhenning Guo,[1] Xiaoyang Yu,[1] Yongkai Deng,[1] Chengyin Wu,[1,2,3] Qihuang Gong,[1,2,3,4] & Yunquan Liu[1,2,3,4,*]

[1] State Key Laboratory for Mesoscopic Physics and Frontiers Science Center for Nano-optoelectronics, School of Physics, Peking University, Beijing 100871, China

[2] Collaborative Innovation Center of Quantum Matter, Beijing 100871, China

[3] Collaborative Innovation Center of Extreme Optics, Shanxi University, Taiyuan, Shanxi 030006, China

[4] Peking University Yangtze Delta Institute of Optoelectronics, Nantong 226010, Jiangsu, China

* Correspondence to: (Y.L.) Yunquan.liu@pku.edu.cn

## Introductory Paragraph

The interaction between a quantum particle's spin angular momentum[1] and its orbital angular momentum[2] is ubiquitous in nature. In optics, the spin-orbit optical phenomenon is closely related with the light-matter interaction[3] and has been of great interest[4,5]. With the development of laser technology[6], the high-power and ultrafast light sources now serve as a crucial tool in revealing the behavior of matters under extreme conditions. The comprehensive knowledge of the spin-orbit interaction for the intense light is of utmost importance. Here, we achieve the in-situ modulation and visualization of the optical orbital-to-spin conversion in strong-field regime. We show that, through manipulating the morphology of femtosecond cylindrical vector vortex pulses[7] by a slit, the photons' orbital angular momentum can be controllably transformed into spin after focusing. By employing strong-field ionization experiment, the orbital-to-spin conversion can be imaged and measured through the photoelectron momentum distributions. Such detection and consequent control of spin-orbit dynamics of intense laser fields have implications on controlling the photoelectron holography and coherent extreme ultraviolet radiation[8].

**Main Text**

In general, light's spin angular momentum (SAM) and orbital angular momentum (OAM) can be independently manipulated and measured in paraxial beams[9]. Whereas a strong coupling between SAM and OAM will occur through interacting with structured materials[10], scattering[11] or focusing[12]. Such spin-orbit interaction (SOI) is inherent in lots of basic optical processes and has attracted increasing attention because of its fundamental and emerging importance, ranging from the optical Hall effects[13], optical image[14], to the control of surface plasmon polaritons[15]. So far, the SOI phenomena mostly involve the spin state affecting on the spatial dynamics of light beams. Recently, some theoretical studies proposed that the inverse process, orbital-to-spin conversion, can be realized through tightly focusing the vortex beams[16,17,18], or scattering the linearly polarized vortex beams by a nanoparticle[19]. Such SOI phenomenon provides a new degree of freedom in optical manipulation, but the experimental verification is still lacking.

With the unprecedented concentration of electromagnetic energy in both time and space, the advances of femtosecond laser technology have significantly facilitated the studies of intense laser-matter interaction[20]. For example, high-order harmonic generation[21] is one of the most important processes as the basis for the table-top coherent extreme ultraviolet source and the attosecond light source. Such a highly nonlinear generation process can be considerably controlled by playing with the photons' spin-orbit state[8]. The SOI phenomenon and the structured laser fields are particularly important in the strong-field community. For a light beam with weak electric field strength, the observation of SOI phenomena can be realized with traditional optical instruments, e.g. using the near-field reconstruction method[22]. However, for an intense laser field, the enormous amounts of energy density can damage the optical instruments in a very short time. Up to now, the SOI phenomenon has been rarely studied and applied in the strong-field regime. How to detect and characterize it for an intense light field is a crucial question.

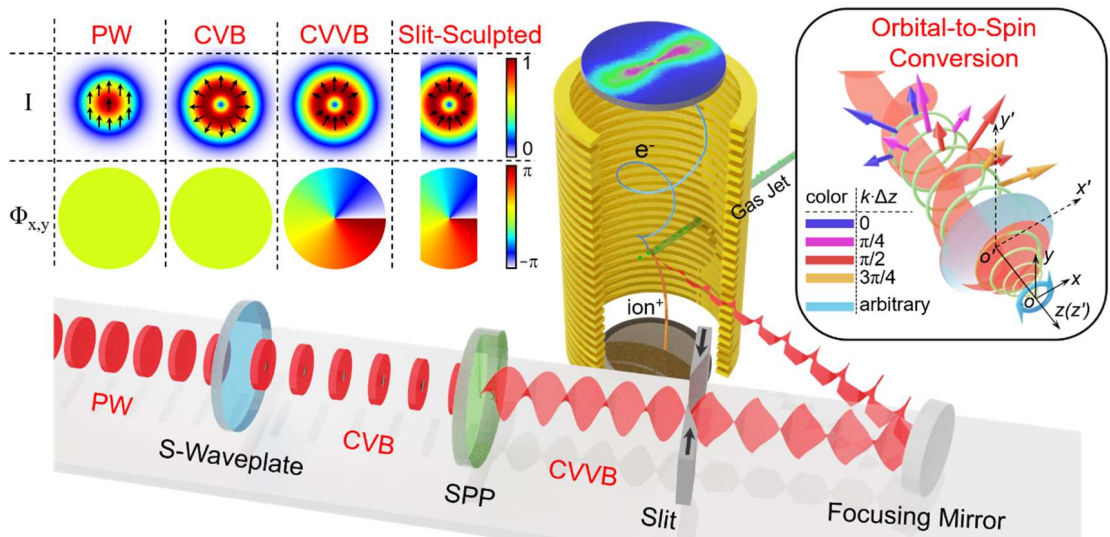

**Figure 1 | Schematic of the experiment.** Layout showing the modulation of light's spatial profile and the detection of strong-field ionization of xenon atoms by COLTRIMS. PW, plane wave; CVB, cylindrical vector beam; CVVB, cylindrical vector vortex beam; S-waveplate, super-structured space-variant waveplate; SPP, spiral phase plate. Left insets: illustrations of light's spatial profile, in which the top panels show the intensity distribution and the instantaneous electric field distribution, and the bottom panels show the structure of the phase of both the x- and y-component electric field. Right inset: illustration of light's orbital-to-spin conversion, in which a lens-based focusing geometry is used for the illustration for the principle. In the right inset, the arrows indicate the locally instantaneous electric field. A circular polarization state is coupled at the focus. The color of the arrows is used to distinguish the phase, $k \cdot \Delta z$, where $k$ is the wavenumber and $\Delta z$ is the relative propagation distance.

Here, we generate the structured femtosecond light field and realize the tunable SOI phenomena in intense light regime at the laser focus. We employ strong-field ionization experiment to detect it in situ. Our experimental scheme is presented in Fig. 1. Utilizing a super-structured space-variant waveplate (S-waveplate) and a spiral phase plate (SPP, here the topological charge is $\ell = 1$), we transform the fundamental plane wave (PW) pulses (790 nm, 25 fs, s-polarization, beam diameter ~11 mm) into cylindrical vector vortex beam (CVVB). In the CVVB's cross-section, the polarization orientations are all along the radial direction. Besides, a screwed phase profile that is characterized by a Hilbert factor $e^{i\ell\varphi}$ ($\varphi$ is the azimuthal angle), is manifested around the optical axis. Hence, the CVVB possesses both the vectorial feature and the spiral phase structure. Then, in order to modulate the orbital-to-spin conversion process as explained later, a horizontal slit with the controllable width is used to tailor the CVVB. The widths of the slit in the experiment are chosen to be $d = $ full-open, 6 mm, 4 mm and 2 mm respectively. After passing through these optical elements, the spatially structured femtosecond light is focused by a 75mm focal length mirror (NA = 0.17). In order to characterize the spin-orbit coupling after the focusing of the beam, a strong-field ionization experiment is carried out. At the focus, the light field interacts with the supersonic xenon atoms gas jet in the high vacuum chamber ($5\times10^{-11}$ mbar) of cold target recoil ion momentum spectroscopy (COLTRIMS) setup[23]. In the interaction region, the peak intensity of the light field was calibrated to be $I \sim 1\times10^{14}$ W / cm$^2$ (the corresponding peak electric field strength is $E \sim 2.5\times10^{10}$ V/cm). The produced photoelectrons are collected by time-of-flight spectrometer with the position-sensitive detector, and their three-dimensional photoelectron momentum distributions are then constructed (see Methods for more details).

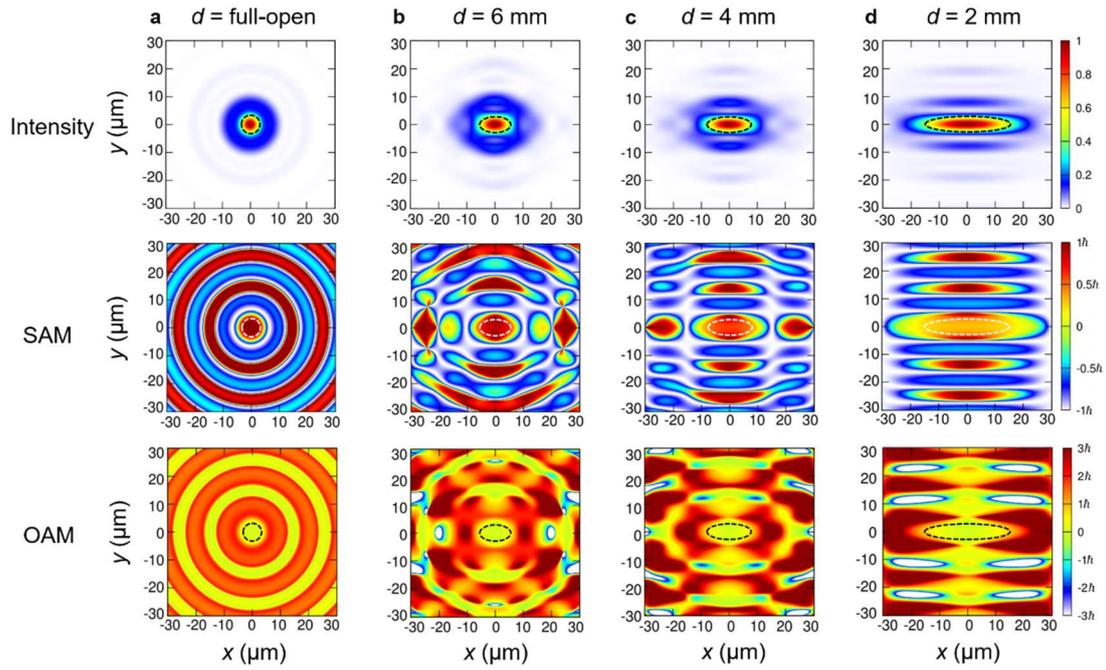

**Figure 2 | Simulations of the longitudinal component of light's intensity, local SAM and OAM distribution in the focal plane. a-d**, The distributions for the different widths of the slit, i.e. full-open, 6 mm, 4 mm and 2 mm respectively. The three panels show the normalized intensity distribution (top), the photon's local SAM distribution (middle) and photon's local OAM distribution (bottom) respectively. The circles in each picture are used to illustrate the full-width-at-half-maximum area of the light spot.

For the light field before focusing, the intensity distribution is a donut-shaped pattern. Notably, as the polarization state at each point of the field is linear and the topological charge of the laser phase is $\ell = 1$, the photon's SAM and OAM is 0 and $\hbar$ respectively. [In this work, we only consider the intrinsic OAM of the light beam (see Methods for more details)]. Here, we employ the Richards-Wolf vectorial theory to analyze the focal light field[24]. One can notice that the intensity distribution of the field reveals a Gaussian-like distribution at the focal plane when the slit is fully opened (Fig. 2a). In the strong-field regime, the optical interaction with atoms or molecules requires a proper description of the local angular momentum per photon. Hence, we further calculate the corresponding distributions of photon's local SAM and local OAM[25] (see Methods for more details). Here, the local SAM and local OAM indicate the SAM and OAM per photon at a given position of the light field. It is worth noting that in our configuration, the numerical aperture is small, and transverse components of the field's angular momentum at the focus are negligible. Thus, we focus on the longitudinal components of the photon's angular momentum. As shown in Fig. 2a, the distribution of the longitudinal local SAM and local OAM are radially varied, exhibiting the concentric-circle pattern. The evident localized orbital-to-spin conversion will show up at the focus. At the center, a typical C-point polarization singularity[26] will be generated, in which the field is circularly polarized.

In this geometry, the mechanism of the orbital-to-spin conversion can be attributed to the directional and time-varying polarization distribution in the incident light's wavefront, as illustrated in the right inset in Fig. 1. At each moment, the spatial maximum of the electric field is oriented to a certain direction (for example, it is upward in the illustration of the CVVB in Fig. 1). And, this direction varies with the laser phase. As known, the electric field at the center of focus is the superposition of the partial waves of the incident field with different positions. Hence, the direction of the electric field in the center of focus will rotate with the varying of laser phase, manifesting the circular polarization ($|SAM| = \hbar$). Based on this, we propose a scheme to control the coupled focal electric field. Because the light field is radially polarized before focusing, if the beam is intercepted by a horizontal slit, the horizontal polarization components of the light field will be curtailed. That is, when the spatial maximum of the incoming electric field rotates to the horizontal direction, the coupled electric field in the focus center will be smaller than which rotates to the vertical direction. Consequently, the electric field at the focus will manifest as the elliptical polarization ($0 < |SAM| < \hbar$).

Here, we control the orbital-to-spin conversion by manipulating the width of the on-path slit. We present the scheme by the calculation of the focal field distributions for the different slit widths, $d$ = 6 mm, 4 mm and 2 mm (Fig. 2b-d), respectively. As shown, when narrowing the slit, the distributions of local SAM and local OAM are evidently varied. We can notice that the distribution evolves regularly in the full-width-at-half-maximum (FWHM) regions of the light spot, which is circled by the dashed lines. Here, the FWHM regions are where the light's intensity is larger than the half maximum of the peak intensity. As shown in Fig. 2, the local OAM is nearly zero at the FWHM region of the light field for each width of the slit. As for the local SAM, it decreases when narrowing the slit. In the strong-field regime, because of the highly nonlinear response, the intense-light-matter interaction process is determined by the FWHM region in the laser field. Thus, realizing the control over the photons' spin-orbit state in this region is particularly important.

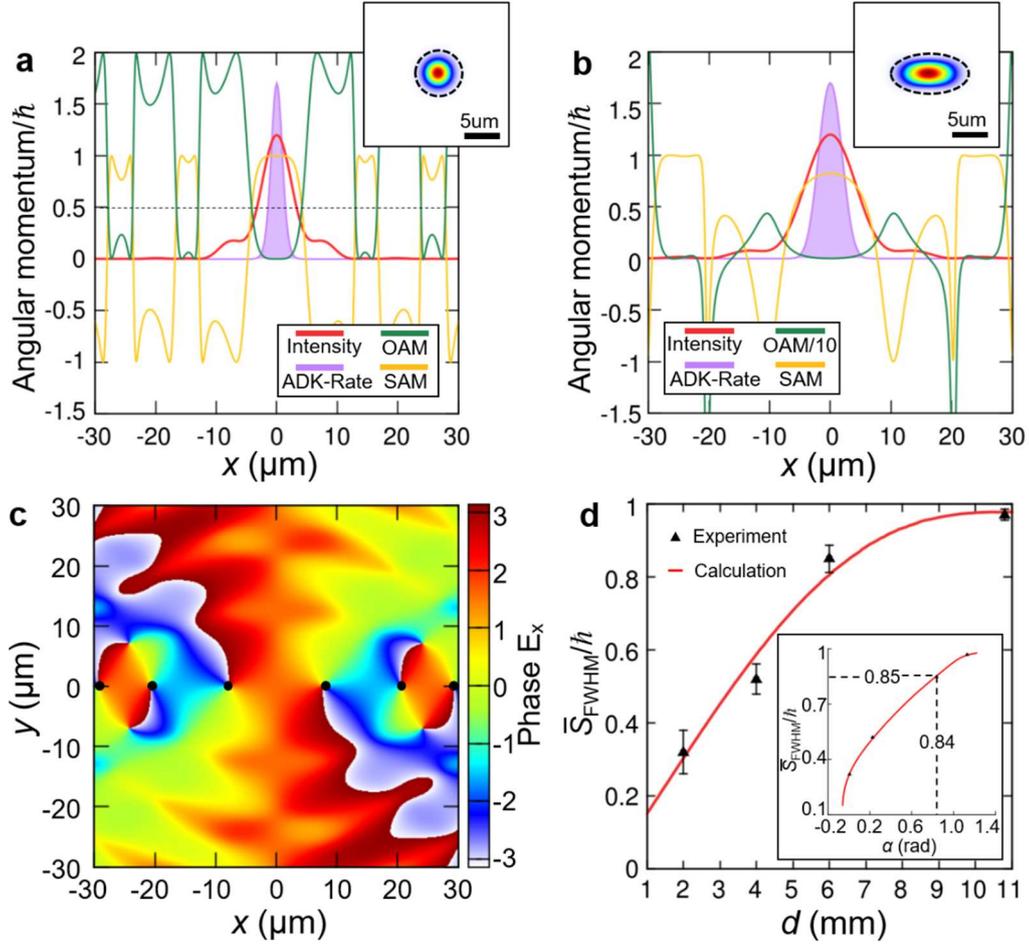

**Figure 3 | Correlation between orbital-to-spin conversion and strong-field ionization. a,** The distribution of light's intensity, photon's local SAM, local OAM and the ionization rate of xenon atoms for the full-open slit along the line, $y = 0$, in the focal plane. The magnitudes of light's intensity and the ionization rate are arbitrary. **b,** Same simulations for the width $d = 6$ mm. Here, the values of local OAM in **b** is divided by 10. The insets in **a** and **b** show the complete distributions of the ionization rate in the focal plane, in which the FWHM region is circled by the line. **c,** The phase distribution of the *x*-component electric field for the slit of $d = 6$ mm. **d,** The FWHM-SAM, $\bar{S}_{FWHM}$, as a function of the slit width. In **d**, the inset demonstrates the retrieval process of experimental conversion efficiencies, in which $α$ is the most probable emitting direction of photoelectrons (See the text for details). The error bars show statistical errors.

Since there is no additional angular momentum adding on the light beam, its total angular momentum (TAM, i.e. the sum of SAM and OAM) is conserved during focusing. Moreover, we calculate the photon's total SAM and OAM, i.e. the spatially weighted averages of the local SAM and OAM, under the paraxial condition. The numerical results indicate that the total SAM and total OAM of the light field are both independently conserved for different slits (see Supplementary Information). Thus, the essence of the orbital-to-spin conversion is actually the spatial redistribution of the photon's SAM and OAM. Then, we can naturally divide the spatial redistribution into two different forms: (1) the SAM and OAM are exchanged only locally, which leads to the consequence that the photon's TAM is spatially homogeneous and equals to the photon's initial TAM ($\hbar$), corresponding to the case of the full-open slit as shown in Fig. 3a; (2) the photon's TAM is transferred between the different positions of the light field, so that the photon's TAM can be spatially varied, as shown in Fig. 3b, in which the slit is $d = 6$ mm. Here, Fig. 3a and Fig. 3b show the photon's angular momentum distributions along the line, $y = 0$, in the focal plane.

For convenience, we call above two different forms as the local and non-local angular momentum transfers, respectively. For the local angular momentum transfer, it reveals the spatially homogeneous TAM distribution, which is found to be closely related with the rotational symmetry of the incident field's amplitude (see Supplementary Information for the details). When controlling the slit to shape the incident beam, the non-local angular momentum transfer will occur and adjust the spatial distribution of the photon's TAM. To better understand this process, we calculate the phase structure of the *x*-component electric field for the slit of $d = 6$ mm (Fig. 3c). As shown, there are several phase singularities (marked with black dots) whose topological charges are $\ell = 1$. If comparing with the full-open slit (see

Supplementary Information), when the slit is $d = 6$ mm, the phase singularities are shifted and the field around the singularities is evidently distorted. One can notice that the locations ($x = \pm 10$ μm, $\pm 20$ μm and $\pm 30$ μm) of these phase singularities coincide with the peaks of the distributions of local SAM and local OAM (Fig. 3b). The shift and distortion of the phase singularities induce the non-local angular momentum transfer in the focal light field (see Supplementary Information for the details).

Though the TAM of the light field is unchanged during focusing, in strong-field physics, we don't need to know or control the entire light field. Instead, the control over the intense region of the laser field is particularly important. To quantify the regional orbital-to-spin conversion, we can define an FWHM-SAM of the focal light field as:

$$\bar{S}_{\text{FWHM}} = \frac{\iint_D S(\mathbf{r})I(\mathbf{r})H[I(\mathbf{r})-I_m/2]d\mathbf{r}}{\iint_D I(\mathbf{r})H[I(\mathbf{r})-I_m/2]d\mathbf{r}}, \quad (1)$$

where $I(\mathbf{r})$ is the local intensity, $S(\mathbf{r})$ is the local SAM, $I_m$ is the peak intensity of the focal field and $D$ represents the focal plane. Note that the integral is confined in the FWHM region of the focal light field by the Heaviside function, $H(t) = \begin{cases} 1, & t \geq 0 \\ 0, & t < 0 \end{cases}$.

Here, the FWHM-SAM can be well understood as the intensity-averaged SAM per photon in the FWHM region of the focal light field.

In Fig. 3d (the solid line), we show the calculated FWHM-SAM as a function of the slit width. As seen, the FWHM-SAM can be widely controlled by modulating the slit. The FWHM-SAM can also qualitatively reflect the strength of the local and non-local angular momentum transfers in the FWHM region of the light field. In general, the strength of the local angular momentum conversion is positively associated with the FWHM-SAM, and the non-local angular momentum conversion is negatively related to the FWHM-SAM (see Supplementary Information).

Although the SOI plays an important role in the optical systems, its effect in such a high-intensity regime is not easy to be measured. Here, we employ an in-situ strong-field ionization method to realize the measurement. To correlate the spatial features of the structured light field with strong-field ionization signals, we have calculated the ionization rate for the model xenon atom ($I_p$ = 12.1 eV). The ionization rate is given by the Ammosov-Delone-Krainov (ADK) theory[27] (see Methods for more details). The ionization rate along with the line, $y = 0$, is shown with the purple curves in Figs. 3a and 3b. One can notice that the ionization rate is sensitive to the light intensity, only the intense region of the light field has a significant effect on the ionization rate. As for the contribution of the weak areas of the light field, it can be naturally filtered by the ionization process. Besides, we have presented the two-dimensional distributions of the ionization rate in the focal plane in the insets of Figs. 3a and 3b, in which the FWHM regions are circled by the dashed lines. As shown, the ionization signals are concentrated in the FWHM region of the light field. It also implies that, by employing photoelectron spectroscopy, what is measured is the localized information of the structured light field. In general, strong-field ionization measurement is expected to be a suitable method to indirectly reflect the localized optical orbital-to-spin conversion of the intense light fields.

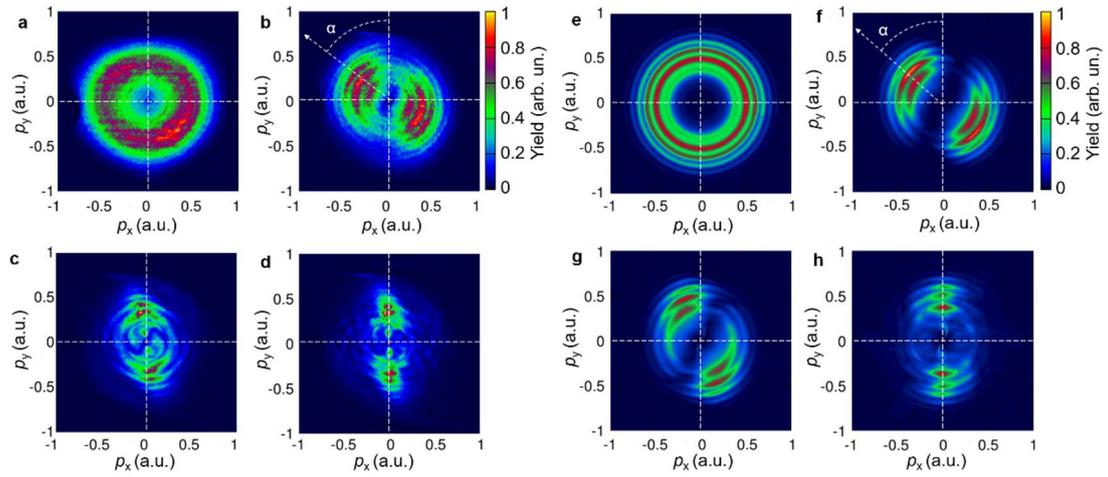

**Figure 4 | The measured and simulated photoelectron momentum distributions. a-d,** The measured photoelectron momentum distributions of xenon atoms ionized by the intense cylindrical vector vortex beam. The widths of the horizontal slit are full-open, 6 mm, 4 mm and 2 mm from **a** to **d**. **e-h,** The simulated photoelectron momentum distributions. The parameters are corresponding to **a-d** respectively.

Here, we experimentally measured the photoelectron momentum distributions of xenon atoms ionized by the structured CVVBs using COLTRIMS[23]. The photoelectron momentum distributions are notably distinct for different widths of the slit (Figs. 4a-d). When the slit is fully opened, the photoelectron momentum distribution is nearly an annulus, which indicates that the electric field is dominated by a circularly polarized state[28]. As for the slit that is 6 mm, the photoelectron momentum distribution is dominated by an elliptically polarized state, manifesting the feature of Coulomb asymmetry[29]. With the further narrowing of the slit, the photoelectrons are mostly populated along the $p_y$ direction. The photoionization is obviously dominated by the linear polarization states and the photoelectrons are mostly distributed along with the field polarization along the *y*-direction.

It should be stressed that the incident laser was prepared with the only OAM but without SAM. Because of the large difference between electron wavepacket (~10 nm) and light's transverse structure (~10 μm) in spatial scales, the light's OAM has less effect on the strong-field ionization process. If there is no orbital-to-spin conversion, the photoelectron momentum distributions would only manifest as that in linearly polarized fields. Hence, the experimental results clearly reveal the orbital-to-spin conversion of the light field. It is worth mentioning that the strong focusing conditions were used in the reported theoretical predictions of the optical orbital-to-spin conversion (e.g. the numerical aperture is NA = 1.26 in Ref. 17). In our experiment, we find that a strong focusing condition is not necessary, which avoids the use of a high-numerical-aperture lens. It can largely facilitate the application in intense femtosecond laser pulses.

To interpret the experimental results, we perform the simulation with the improved quantum-trajectory Monte Carlo (QTMC) model[30]. In the model, we have fully considered the spatial structure of the focal electric field. The details of the theoretical model are summarized in the Methods. As shown in Figs. 4e-h, the simulation results are in good agreement with the measurements. The photoelectron momentum distributions reveal the characteristics of photoionization by circularly, elliptically, and linearly polarized state. For each photoelectron momentum distribution, it is contributed by miscellaneous polarization states in the focal plane. The macroscopic appearance of the photoelectrons is mainly contributed by the polarization states of the dominated electric field. As shown, the photoelectron momentum distributions are dependent on the morphology of the incoming light beam. Such an interesting phenomenon contains the intrinsic relationship between the optical angular momentum transition and the strong-field ionization process.

Based on the simulation, we can extract the aforementioned FWHM-SAM, $\bar{S}_{\text{FWHM}}$. First of all, it is necessary to define a representative variable, which can bridge the photoelectron momentum distributions and the FWHM-SAM of the focal light field. Here, we choose the deflection angle of the most probable emitting direction of photoelectrons, as labeled by $\alpha$ in Fig. 4b and Fig. 4f. This variable is monotonically correlated with the FWHM-SAM. In principle, the tilted angle is dependent on the joint effect of the light field feature and the photoionization dynamics.

We calculate the FWHM-SAM $\bar{S}_{\text{FWHM}}$ and the corresponding deflection angle $\alpha$ for different slit widths ($d$ = 1 mm, 1.2 mm, …, 11 mm). And then, we use cubic spline interpolation to numerically establish their correlation $\bar{S}_{\text{FWHM}}(\alpha)$, as shown in the inset of Fig. 3d. By comparing the measured deflection angles with the theoretical values, we can retrieve the FWHM-SAM of the focal light field in the experiment. For example, when the width of the slit is $d$ = 6 mm, the deflection angle is $\alpha$ = 0.84 rad.

According to the correlation $\bar{S}_{\text{FWHM}}(\alpha)$, we can obtain the FWHM-SAM $\sim 0.85\hbar$. Likewise, for $d$ = full-open, 4 mm and 2 mm, the FWHM-SAMs are $0.97\hbar$, $0.52\hbar$, and $0.32\hbar$ respectively. In general, the extracted FWHM-SAMs match the theoretical predictions well (Fig. 3d). Here, a slight deviation is visible. This might be predominantly caused by the propagation effect of the structured light pulses between the slit and the focusing mirror. This is because, as we demonstrate in this work, the orbital-to-spin conversion is significantly controlled by the morphology of the light beam. When propagating the light field, its morphology, more or less, will be changed, therefore leading to the change of the orbital-to-spin conversion.

In conclusion, we have realized the in-situ detection of the spin-orbit interaction of intense femtosecond laser pulses based on strong-field ionization experiment. Through modifying the morphology of the incident cylindrical vector vortex beam, a controllable orbital-to-spin conversion scheme has been demonstrated. This work reveals the non-negligible effect of light's spatial structure in focusing for intense laser fields, which provides for a deep insight into the interaction of intense light fields with matters. The proposed optical spin-orbit interaction scheme provides a new degree of freedom in optical manipulation, that can be employed to control the extreme ultraviolet radiation using structured intense light fields. We also show that the strong-field ionization has the potential to characterize complex structured intense light fields through photoelectron mapping. Based on this work, the detection of the intense light beams' topology, such as in the form of polarization Möbius strips[31], can be expected.

## Methods

**Light modulation.** The s-polarization laser pulses are delivered from a multipass Ti:sapphire laser amplifier at a central wavelength of 790 nm. The repetition rate of the laser pulse is 3 kHz and the pulse duration is 25 fs. The beam diameter is ~11 mm (full-width at half-maximum in intensity). The intensity and polarization of the light field are controlled by wire grid polarizers, λ/2 and λ/4 retardation plates, respectively. A commercial super-structured space-variant waveplate[32] is used to structure the fundamental plane wave to a cylindrical vector beam. We control the fast-axis orientation of the waveplate along the direction of the laser polarization so that the generated light beam is radially polarized. The phase structure of the light beam is imprinted by a commercial spiral phase plate[33] whose center wavelength is 790 nm and the topological charge is 1. The home-made horizontal slit is controlled by a one-dimensional linear stage. These optical elements are arrayed as close as possible. To make sure that the polarization singularity, phase singularity, and optical axis are well superimposed, every optical element is mounted on a two-dimensional motorized linear stage, whose minimum achievable incremental movement is 0.05 μm. The light beam is focused by a silver-coated concave mirror ($f$ = 75 mm), which is placed inside the high vacuum chamber of COLTRIMS.

**COLTRIMS technique**. The supersonic gas jet of xenon atoms is delivered along *x*-direction by a small nozzle with the opening hole diameter 30 μm. Subsequently, the gas jet is filtered by a skimmer (diameter 290 μm) and then collimated by a pair of small holes (diameter 1 mm). The transversal size of the gas jet is approximately equal to the size of the holes. The focus volume of the light beam is much smaller than the size of the gas jet. Electrons and ions are produced by the strong-field ionization in the light focus. The released particles are guided to the two-dimensional position-sensitive detectors by a homogeneous electric field (~3.3 V/cm) along the time-of-flight (TOF) axis. The length of the field acceleration region is 7.02 cm. For the electrons, the acceleration region is followed by a field-free region with a length of 14 cm. The Helmholtz coils supply a weak homogeneous magnetic field (~6.5 Gauss) to compensate the earth magnetic field and the spectrometer is further shielded from remaining magnetic fields by a μ-metal shield. The signal of the electrons (ions) is amplified by a stack of 3 (2) multi-channel plates with a diameter of 77 mm (80 mm). For electrons (ions), a delay-line anode with three (two) layers is used to measure the three-dimensional momentum. The spectrometer and detector work under ultra-high vacuum ($5\times10^{-11}$ mbar). The electrons and their parent ions are measured in coincidence to avoid the false signals of electrons. The signals from the delay-line detectors are amplified through the FAMP8 module and then transformed into "digital" NIM signal with a Constant Fraction Discriminator circuit. We use a counter to monitor the rate of ionization events. In the experiment, when changing the width of the slit, the electric field strength at the focus is inevitably varied. In the experiment, we control the power of the incident light beam to keep the counter rate at ~1000/s for different slits. Here, the peak electric field strength when the slit was fully opened is accurately calibrated by performing the intensity-continuously-varying experiments[34]. In the data processing, we define an opening angle $\beta_z$, which is the angle between the propagation direction of the light beam and the emitting direction of the photoelectron. To improve the resolution of the photoelectron momentum distributions, the photoelectrons whose momentum is beyond the constraint, $60° < \beta_z < 120°$, are abandoned in the presented photoelectron momentum distributions (Fig. 4a-d).

**Light field simulation.** The coordinates are illustrated in the right inset of Fig. 1. As shown, to simplify the mathematical expression of the light field, the origins of the coordinates for the incident light field and the focal light field are chosen to be different. For these two light fields, the origins are placed on the center of the incident beam's cross-section ($O'$) and the center of the focal light field ($O$), respectively. To differentiate the variables of the incident light field from that of the focused light field, the former is marked by the superscript $'$.

In spite of the temporal envelope of the laser pulse, the electric field distribution of the incident light beam can be expressed as:

$$\mathbf{E}'(\rho',\varphi') = F(\rho',\varphi')e^{i\ell\varphi'}H(d/2-|\rho'\cos\varphi'|)\mathbf{e}_{\rho'}, \qquad (M1)$$

where $\rho' = (x'^2 + y'^2)^{1/2}$ is the radial distance in the incident plane; $\varphi' = \mathrm{atan}(y'/x')$ is the azimuthal angle in the incident plane; $F(\rho',\varphi')$ is the amplitude distribution in the polarization plane; the term $e^{i\ell\varphi'}$ represents the helical phase with the topological charge $\ell$; the effect of the slit is described by the Heaviside function, $H(d/2-|\rho'\cos\varphi'|)$; $\mathbf{e}_{\rho'}$ represents the radial unit vector.

To accurately reproduce the electric field at the focal plane, we employ the Richards-Wolf vectorial diffraction method[24] to calculate the field near the focus at $\mathbf{r}=(\rho,\varphi,z=0)$:

$$\mathbf{E}(\mathbf{r}) = \frac{-ikf}{2\pi} \int_0^{\theta'_m}\int_0^{2\pi} F(\rho',\varphi')H(d/2-|\rho'\cos\varphi'|)P(\theta')$$
$$\times \exp(i\ell\varphi')\cdot\exp(ik[z\cos\theta' + \rho\sin\theta'\cos(\varphi'-\varphi)]) \qquad (M2)$$
$$\times\left[\xi'\begin{pmatrix}\cos\theta'\cos(\varphi'-\varphi)\mathbf{e}_\rho \\ \cos\theta'\sin(\varphi'-\varphi)\mathbf{e}_\varphi \\ -\sin\theta'\mathbf{e}_z\end{pmatrix} + \gamma'\begin{pmatrix}-\sin(\varphi'-\varphi)\mathbf{e}_\rho \\ \cos(\varphi'-\varphi)\mathbf{e}_\varphi \\ 0\end{pmatrix}\right]\sin\theta'd\varphi'd\theta',$$

where $\rho$ and $\varphi$ are the radial distance and the azimuthal angle in the focal plane respectively, $\xi'$ and $\gamma'$ are the radial and azimuthal amplitude factors respectively, $k = 2\pi/\lambda$ is the wavenumber, $\theta'$ is the polar angle in the output pupil of the focusing system, $\theta'_m$ is the maximal angle determined by the numerical aperture and $P(\theta') = \sqrt{\cos\theta'}$ is the apodization factor. Geometrically, we have $\rho' = f\tan\theta'$, and $f$ is the focal length of the mirror. In the simulations, the radially polarized field is used, and thus the radial and azimuthal factors are $\xi' = 1$ and $\gamma' = 0$, respectively. For the sake of simplicity, a monochrome laser field with $\lambda = 790$ nm is considered. And, the spatial intensity distribution of the incident field is approximated to be $F(\rho',\varphi') = 1$ when $\rho' \leq 5.5$ mm and $F(\rho',\varphi') = 0$ when $\rho' > 5.5$ mm. During the numerical calculation, the 20th order Gauss-Legendre integral formula is used. The numerical grid size is 80 μm in the focal plane with a spacing of $\Delta x = 0.4$ μm.

Based on the calculated electric field components in the focal region, the angular momentum per photon **J(r)** at the focal plane about the optical axis **r** = 0 can be calculated[35,36]. Adopting the Minkowski form[37], the angular momentum density of the light field can be expressed as $\mathbf{AM}(\mathbf{r}) = \text{Re}\left[\mathbf{r} \times (\mathbf{D}^*(\mathbf{r}) \times \mathbf{B}(\mathbf{r}))\right]$. Here, the standard notations are used and the superscript * denotes the complex conjugate. The energy density of the light field can be given by $w(\mathbf{r}) = \frac{1}{2}\text{Re}\left[\mathbf{E}^*(\mathbf{r}) \cdot \mathbf{D}(\mathbf{r}) + \mathbf{B}^*(\mathbf{r}) \cdot \mathbf{H}(\mathbf{r})\right]$. With these, the local angular momentum value per photon at a given position can be written as:

$$\mathbf{J}(\mathbf{r}) = \hbar\omega \frac{\mathbf{AM}(\mathbf{r})}{w(\mathbf{r})}. \tag{M3}$$

Here, $\omega$ is the angular frequency of the photon. In isotropous media, the expression of the angular momentum can be simplified by Maxwell's equations and the constitutive relations, with the result:

$$\mathbf{J}(\mathbf{r}) = \hbar \frac{2\,\text{Im}\left[\mathbf{r} \times (\mathbf{E}^*(\mathbf{r}) \times (\nabla \times \mathbf{E}(\mathbf{r})))\right]}{\text{Re}\left[\mathbf{E}^*(\mathbf{r}) \cdot \mathbf{E}(\mathbf{r}) + \frac{1}{k^2}(\nabla \times \mathbf{E}^*(\mathbf{r})) \cdot (\nabla \times \mathbf{E}(\mathbf{r}))\right]}. \quad (M4)$$

Under the paraxial approximation, the denominator in the above equation can be further simplified: $\mathbf{E}^*(\mathbf{r}) \cdot \mathbf{E}(\mathbf{r}) \approx \frac{1}{k^2}(\nabla \times \mathbf{E}^*(\mathbf{r})) \cdot (\nabla \times \mathbf{E}(\mathbf{r}))$. In the experiment, since the focal length ($f$ = 75 mm) is much larger than the beam diameter (~11 mm), the longitudinal electric field is trivial. For example, when the slit is full-open, the peak electric field strength of the $x$ and $y$ component electric fields are ~400 times larger than the $z$ component. Then, the transverse angular momentum can be neglected and we focus on the longitudinal angular momentum:

$$J_z(\mathbf{r}) = \hbar \frac{\text{Im}\left\{\mathbf{E}^* \cdot \left[\mathbf{e}_z \cdot (\mathbf{r} \times (-i\nabla))\right]\mathbf{E} + \mathbf{e}_z \cdot \mathbf{E}^* \times \mathbf{E}\right\}}{\text{Re}\left[\mathbf{E}^*(\mathbf{r}) \cdot \mathbf{E}(\mathbf{r})\right]}. \quad (M5)$$

This equation is suitable for calculating the photon's angular momentum of the vector beams. Notably, this equation contains two terms. For paraxial fields, the first term is the photon's orbital angular momentum and the second term is the photon's spin angular momentum.

To be exact, the OAM of a light field can be divided into two types, including the intrinsic OAM[2] and the extrinsic OAM[38]. The intrinsic OAM is related to the optical vortices inside the beam, and the extrinsic OAM is associated with the beam trajectory. One should note that in this work, the mentioned "OAM" only refers to the intrinsic OAM of the light beam. And the "orbital-to-spin conversion" refers to the conversion between the intrinsic OAM and the SAM of the light field.

**Photoionization simulation.** The cycle-averaged photoionization ADK rate (Fig. 3a and 3b) of a monochromatic light is given by:

$$W_{ADK} = A(\varepsilon, E) \frac{ED^2}{8\pi Z} \exp(-\frac{2Z^3}{3n^{*3}E}), \quad (M6)$$

where $n^* = Z/\sqrt{2I_p}$, $D = (4Z^3/En^{*4})^{n^*}$, $Z$ is the nuclear charge and $I_p$ is the ionization potential, and $A(\varepsilon, E)$ is the cycle-averaged coefficient. When the light is an elliptical field with the ellipticity $\varepsilon$, the cycle-averaged coefficient can be expressed as[39]:

$$A(\varepsilon, E) = \left[\frac{\varepsilon(1+\varepsilon)}{2}\right]^{-1/2} a\left(\frac{1-\varepsilon}{3\varepsilon} \frac{Z^3}{En^{*3}}\right), \quad (M7)$$

in which, $a(x) = e^{-x}I_0(x)$ is a monotonically decreasing function, and $I_0(x)$ is the Bessel function of imaginary argument. And it has: $a(0)=1$, $a(x) \sim (2\pi x)^{-1/2}$ for $x \gg 1$.

On the other hand, to simulate the photoelectron momentum distributions of xenon atoms in the spatially structured light fields, we improve the quantum-trajectory Monte Carlo (QTMC) model[30] based on the calculation of the electric field distribution in the focal plane, as described in the 'Light field simulation' section. For the calculation of the electric field at the focal plane, a 200×200 calculation grid with the bin size of $\Delta x$ = 0.4 μm is selected. Within each bin, the electric field is treated as a plane wave with a unified polarization state. Since the de Broglie wavelength of the photoelectrons is much smaller than the size of the light spot, the interference of photoelectrons that comes from the same bin is coherently summed up in the calculation. Moreover, because the size of the gas jet is much larger than the light spot in the focus, the weights for each calculation grid are unified. Consequently, the probability in each final momentum bin is given by

$$M_{bin} = \sum_i \left| \sum_j \sqrt{W} \exp(-iQ) \right|^2$$, where $i$ is the label of the calculation grid of the focal light field, $j$ is the label of the electron trajectory in the bin, $W$ is the weight of the trajectory and $Q$ is the classical action. Here, we use an eight-cycles half trapezoidal envelope which is composed of a seven-cycle plateau and a one-cycle ramp-off. The ionization time is confined within the first two cycles. For the final momentum distributions, a 300×300 calculation grid with the bin size of $\Delta p = 0.01$ a.u. is selected. The convergence of our calculations and the stability for different widths of slit have been checked. We should note that, currently, the semi-classical model may be the only and the most efficient approach to calculate the photoelectron momentum distributions of strong-field ionization with the spatially structured field.

## Data availability

The data that support the plots within this paper and other findings of this study are available from the corresponding author upon reasonable request.

## Acknowledgments

We thank the finance support by the National Science Foundation of China (Grant No. 9205020002, 91850111, 11434002, 11774013 and 11527901).

## Author contributions

Y. Fang and Z. Guo performed the experiments. Y. Fang and Y. Liu analyzed and interpreted the data. Simulations were implemented by Y. Fang and M. Han. This project was coordinated by Y. Liu. All authors discussed the results and wrote the paper.

## Competing financial interests

The authors declare no competing financial interests.

## Materials & Correspondence

Correspondence and requests for materials should be addressed to Y. Liu (yunquan.liu@pku.edu.cn).